\documentclass[12pt]{article}

\usepackage{scicite}
\usepackage{times}
\usepackage{cite}
\usepackage{url}
\usepackage{amssymb,amsmath,amsthm}
\usepackage{bm}
\usepackage{graphicx}
\usepackage{enumerate}
\usepackage{microtype}
\usepackage{color}
\usepackage{hyperref}
\usepackage{multirow}

\theoremstyle{plain}
\newtheorem*{theorem*}{Theorem}

\definecolor{webgreen}{rgb}{0,.35,0}
\definecolor{webbrown}{rgb}{.6,0,0}
\definecolor{RoyalBlue}{rgb}{0,0,0.9}
\definecolor{purp}{rgb}{0.6,0.05,0.8}
\definecolor{ora}{rgb}{0.7,0.35,0.02}

\hypersetup{
   colorlinks=true, linktocpage=true, 
   urlcolor=webbrown, linkcolor=RoyalBlue, citecolor=webgreen,
   pdfauthor={Gary P. T. Choi, Lucy Liu, L. Mahadevan},
   pdfsubject={Explosive rigidity percolation in kirigami}
}

\newcounter{lastnote}

\newtheorem{proposition}{\bf Proposition}[section]

\begin{document}
\author{Gary P. T. Choi$^{1}$, Lucy Liu$^{2}$, L. Mahadevan$^{2,3\ast}$\\
\\
\footnotesize{$^{1}$Department of Mathematics, Massachusetts Institute of Technology, Cambridge, MA, USA}\\
\footnotesize{$^{2}$School of Engineering and Applied Sciences, Harvard University, Cambridge, MA, USA}\\
\footnotesize{$^{3}$Departments of Physics, and Organismic and Evolutionary Biology, Harvard University, Cambridge, MA, USA}\\
\footnotesize{$^\ast$To whom correspondence should be addressed; E-mail: lmahadev@g.harvard.edu}
}
\title{Explosive rigidity percolation in kirigami}
\date{} 

\baselineskip24pt

\maketitle

\begin{abstract}
Controlling the connectivity and rigidity of kirigami, i.e. the process of cutting paper to deploy it into an articulated system, is critical in the manifestations of kirigami in art, science and technology, as it provides the resulting metamaterial with a range of mechanical and geometric properties. Here we combine deterministic and stochastic approaches for the control of rigidity in kirigami using the power of $k$ choices, an approach borrowed from the statistical mechanics of explosive percolation transitions. We show that several methods for rigidifying a kirigami system by incrementally changing either the connectivity or the rigidity of individual components allow us to control the nature of the explosive transition by a choice of selection rules. Our results suggest simple lessons for the design and control of mechanical metamaterials.
\end{abstract}

\section{Introduction}
Kirigami, the art of cutting paper to make it morph into different shapes via articulated local rotations while preserving global connectivity has been extensively studied in recent years as a new paradigm for mechanical metamaterials with applications to a range of problems in science and engineering~\cite{callens2018flat,zhai2021mechanical}. Deterministic approaches to kirigami have primarily focused on understanding the geometry and mechanical response of periodic cut patterns, with tiles ranging from simple regular polygons~\cite{grima2000auxetic} to more general wallpaper group patterns~\cite{rafsanjani2016bistable,stavric2019geometrical,liu2021wallpaper}. Complementing this, the design of non-periodic kirigami patterns modulates the geometry~\cite{konakovic2018rapid,choi2019programming,choi2021compact,dang2021theorem,dudte2022additive} and the topology~\cite{lubbers2019excess,chen2020deterministic,liu2022quasicrystal} of the cut patterns to achieve different geometric and mechanical properties.

Since kirigami requires the presence of cuts, a natural question is that of bond and rigidity percolation as a function of the topology of the cuts. Inspired by the question of rigidity percolation in bond networks~\cite{jacobs1996generic,ellenbroek2011rigidity,zhang2015rigidity}, we recently explored the control of the rigidity and connectivity of kirigami patterns by changing the pattern topology, either deterministically or stochastically~\cite{chen2020deterministic}. In particular, we established a constructive theorem for deterministically rigidifying a kirigami pattern using a minimum number of tile connections. Separately, we also explored a stochastic analog of this question and showed that rigidity and connectivity arise via a percolation transition controlled by the fraction of tile connections. Given the rough similarities between the observed connectivity patterns obtained using the deterministic approach with that obtained via the stochastic approach near the phase transition, it is natural to ask whether we can interpolate between the deterministic and the stochastic control of connectivity and rigidity. In addition to being of intrinsic interest, the answer to the question is likely to be of relevance in the context of practical control of the rigidity of mechanical metamaterials. 

\begin{figure}[t!]
    \centering
    \includegraphics[width=0.7\linewidth]{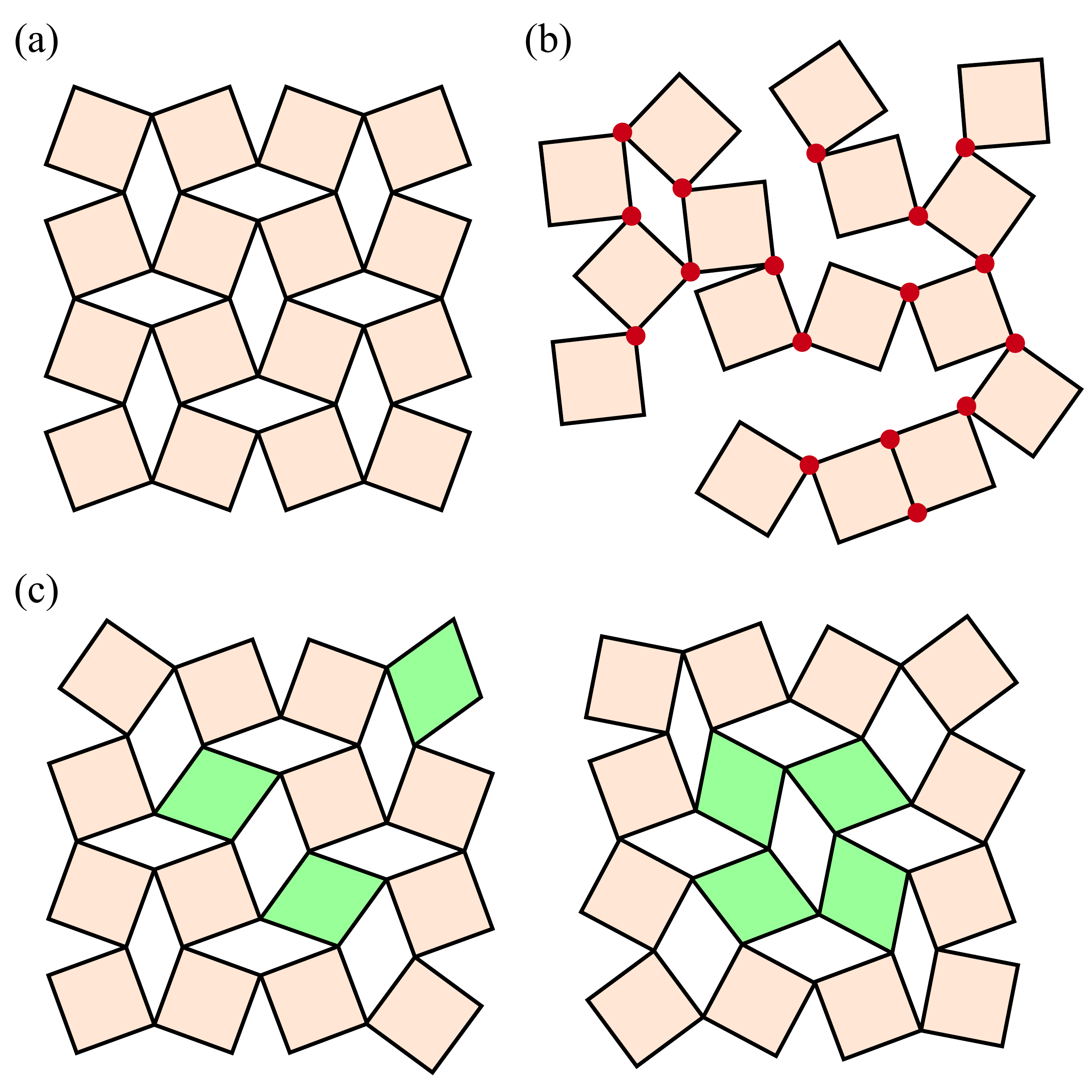}
    \caption{Rigidity control in kirigami. (a) A standard rotating squares pattern, with all tiles being squares connected in an ordered manner. (b) Fixing the rigidity of all tiles and changing the connectivity of them can lead to a change in the rigidity of the overall structure. Here, the tile connections are highlighted in red. (c) Keeping the standard connectivity and changing the rigidity of certain components can also lead to a change in the overall structure rigidity as illustrated by the two examples shown. Here, the tiles that are allowed to be floppy are highlighted in green. It is noteworthy that as shown in the second example, enforcing the rigidity of all boundary tiles does not necessarily constrain the rigidity of the interior tiles. In other words, there can be multiple floppy modes in such patterns.}
    \label{fig:overview}
\end{figure}

In this study, we show how the concept of explosive percolation~\cite{achlioptas2009explosive,radicchi2009explosive,ziff2010scaling,araujo2010explosive,da2010explosive} provides a natural way to parameterize this transition. Explosive percolation was first studied in random networks as a way to tune the nature of the percolation transition by adding/subtracting edges to a network one by one using \emph{the power of two choices}. At the simplest level, this follows from a choice to create/break only one edge randomly at each step by sampling two (or more) random edges and selecting one based on certain selection rules. The set of rules can then be used to qualitatively change the nature of the percolation transition~\cite{d2019explosive}. As we will see, by using different rules for how to connect kirigami tiles, we can naturally move between the deterministic and the stochastic approaches to control the rigidity of the pattern, just as in explosive percolation. To simplify our discussion, we focus on quad kirigami patterns with $L \times L$ quadrilateral tiles (Fig.~\ref{fig:overview}(a)) and consider the following approaches to control the pattern rigidity: 
\begin{enumerate}[(i)]
    \item Fixing the tile rigidity and changing the pattern topology by adding a connection between two tiles in each step (Fig.~\ref{fig:overview}(b)). This connection-based method is described in Section~\ref{sect:connection}.
    \item Fixing the pattern topology and changing the tile rigidity by fixing either the angles or vertex coordinates of a component in each step (Fig.~\ref{fig:overview}(c)). These angle-based and coordinated-based methods are described in Section~\ref{sect:angle} and \ref{sect:coordinate} respectively.
\end{enumerate}
For each approach, we show how \emph{the power of $k$ choices} can lead to significantly different percolation transitions.

\section{Connection-based method}\label{sect:connection}
Consider an $L \times L$ quad kirigami pattern where all $L\times L$ tiles are disconnected and rigid; for most of our numerical experiments, we let $L = 5, 10, 15, 20$. Then, the kirigami pattern has a total degrees of freedom (DOF) of $3L^2$. In every step, we add a connection between two neighboring tiles, creating a linkage. We note that adding a connection is equivalent to enforcing $x_i=x_j$ and $y_i=y_j$ for two nodes $(x_i,y_i)$ and $(x_j,y_j)$, and hence the total DOF decreases by at most 2. In earlier work~\cite{chen2020deterministic}, we have shown that by adding appropriate connections between neighboring tiles, one can rigidify the entire kirigami pattern in the most efficient way, i.e. with the smallest number of links, added deterministically. In contrast, if we add the connections randomly one by one, we observe a linear to sublinear transition in the total DOF of the pattern as the number of connections increases~\cite{chen2020deterministic}, with a percolation transition that signifies the switch in behaviour. 

If instead of a random connection choice,  we consider $k$ random candidate connections every time and choose one among them based on some selection rule, this provides a degree of control over the rigidity percolation and leads to highly unusual effects. Here, we consider the number of choices from $k = 1$ to $k=25$, and for each pair $(L, k)$, we perform 200 simulations to add connections to an $L\times L$ kirigami system gradually. We note that there are in total $4L(L-1)$ possible connections between all neighboring tiles in an $L \times L$ pattern. In each step, we sample $k$ unused connections as the candidates and pick one of them to add based on certain selection rule. We then compute the total DOF of the current pattern using the rank of the infinitesimal rigidity matrix~\cite{guest2006stiffness}, noting that the pattern is rigid if and only if the total DOF is exactly equal to 3, the total number of global DoF for planar systems. The above process is continued until all $4L(L-1)$ connections are added. Denoting $r$ as the percentage of total connections added, we consider the number of rigid patterns among all 200 simulations for different $r$ and quantify the probability of getting a rigid pattern $P \in [0, 1]$ as $r$ increases from $0$ to $1$. We consider two selection rules:
\begin{enumerate}[(1)]
\item \textbf{Most efficient connection rule}: We always choose the connection that gives the minimum total DOF among all $k$ choices, i.e. in the most efficient way of rigidifying the pattern. Using this rule, we can significantly accelerate the rigidity percolation. 
\item \textbf{Least efficient connection rule}: Conversely, if we always choose the connection that gives the maximum total DOF among all $k$ choices, i.e. in the least efficient way of rigidifying the pattern, we can significantly delay the rigidity percolation.
\end{enumerate}

\begin{figure}[t!]
    \centering
    \includegraphics[width=0.75\linewidth]{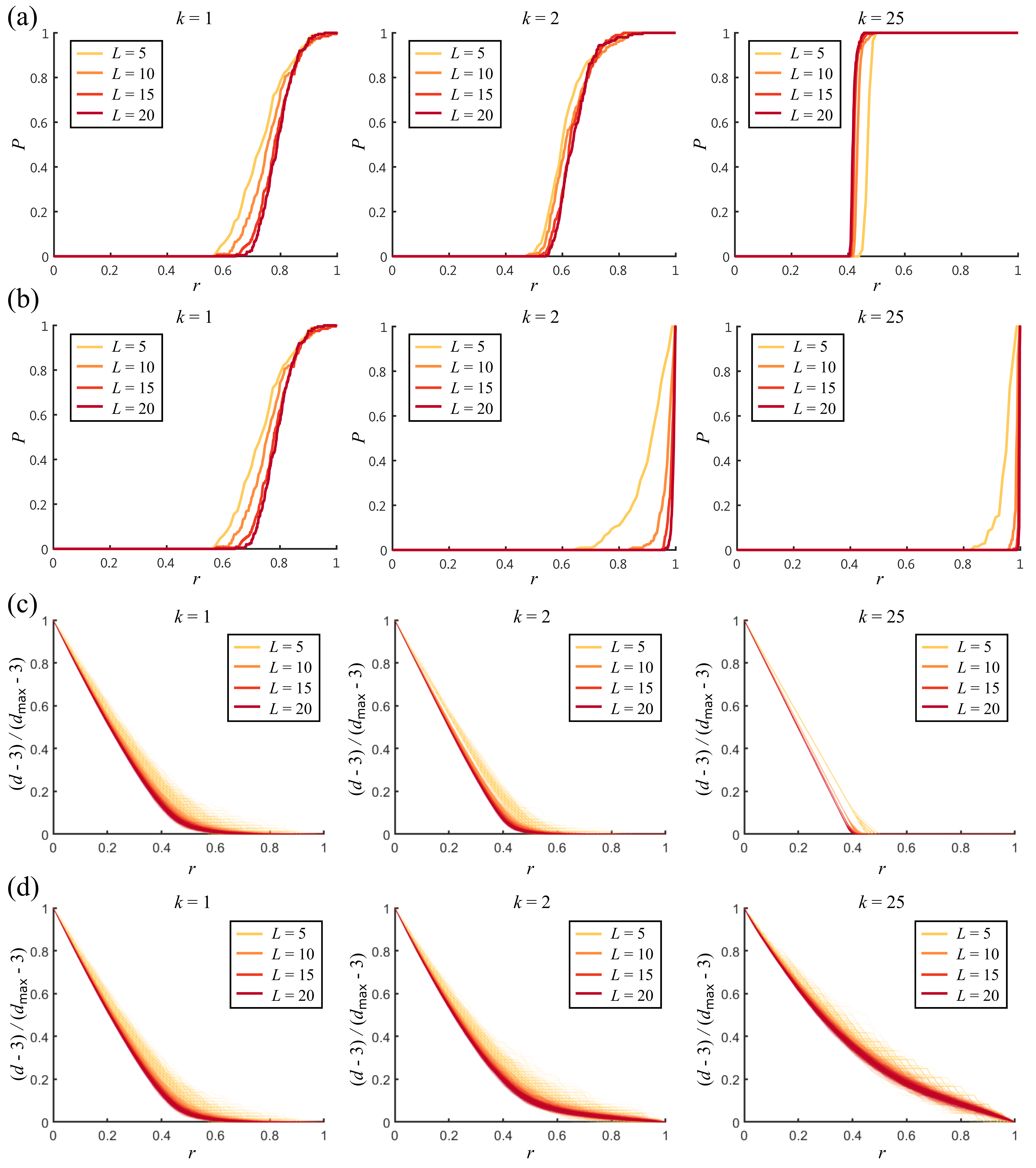}
    \caption{Explosive rigidity percolation achieved using the connection-based method. (a) Using the power of $k$ choices and the most efficient connection rule, we can accelerate the rigidity percolation, with a very sharp transition in the probability of getting a rigid pattern $P$ from 0 to 1 at around $r = 0.4$, where $r$ is the percentage of connections added. (b) Using the power of $k$ choices and the least efficient connection rule, we can delay the rigidity percolation and achieve a very sharp transition in $P$ from 0 to 1 at around $r = 1$. (c) The normalised total degrees of freedom obtained using the most efficient connection rule for all 200 simulations for each $L$, where each curve represents one simulation, $d$ is the total DOF and $d_{\text{max}} = 3L^2$ is the maximum DOF in an $L\times L$ pattern. (d) The normalised total degrees of freedom obtained using the least efficient connection rule for all 200 simulations for each $L$.}
    \label{fig:link_result}
\end{figure}

As shown in Fig.~\ref{fig:link_result}(a), by using the most efficient connection rule and increasing $k$ from $1$ (corresponding to a single random choice) all the way to 25, we note that the transition associated with the rapid increase in the probability of getting a rigid pattern becomes sharper and also occurs earlier at around $r = 0.4$. By contrast, using the least efficient connection rule, a sharp transition in $P$ from 0 to 1 occurs much later at around $r = 1$ (see Fig.~\ref{fig:link_result}(b)). In this latter case, the patterns are not rigid until almost all possible connections are added. We can also consider the change in the normalised total DOF $(d-3)/(3L^2-3)$ for each simulation, where $d$ is the total DOF of the pattern. For the most efficient connection rule, we can see in Fig.~\ref{fig:link_result}(c) that the linear-sublinear transition in the normalised DOF becomes very sharp and all patterns become rigid at around $r = 0.4$ for large $k$. By contrast, the least efficient connection rule significantly slows down the transition in the normalised DOF and none of the patterns are rigid until $r$ is close to 1 (Fig.~\ref{fig:link_result}(d)).

\begin{figure}[t!]
    \centering
    \includegraphics[width=0.85\linewidth]{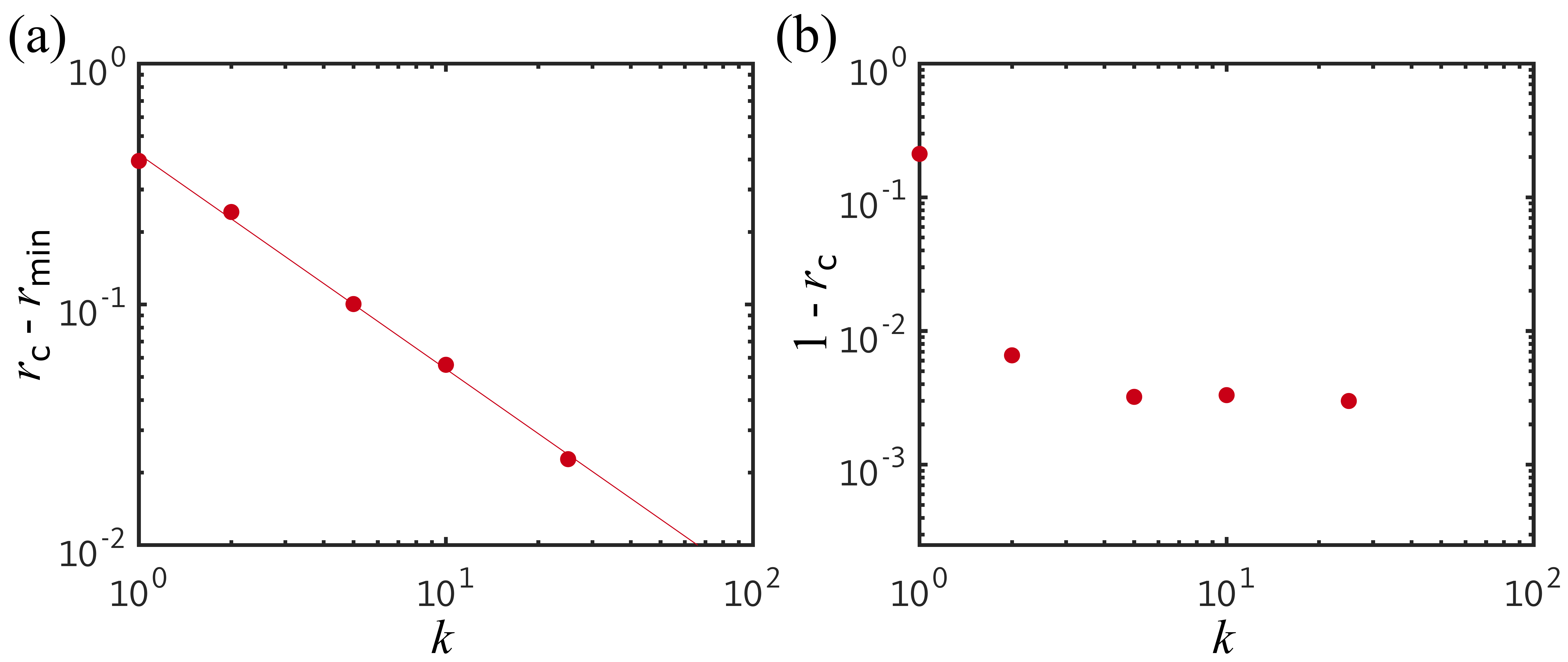}
    \caption{Analyses of different selection rules for the connection-based method. (a) A log-log plot of the number of choices $k$ and the difference $r_c - r_{\text{min}}$ for the most efficient connection rule, with $L = 20$. Each dot corresponds to one choice of $k$ ($k = 1, 2, 5, 10, 25$), and the red line is the best-fit straight line. (b) A log-log plot of the number of choices $k$ and the difference $1 - r_c$ for the least efficient connection rule, with $L = 20$.}
    \label{fig:link_analysis}
\end{figure}

Moving towards a more quantitative analysis of the connection-based method, we define the critical percentage of connections $r_c$ as the value of $r$ with $P = 1/2$ and study how $r_c$ changes with $k$ for the two selection rules. For the most efficient connection rule, it was proved in~\cite{chen2020deterministic} that for the deterministic case, the minimum number of steps needed for rigidifying an $L\times L$ kirigami (where $L \geq 2$) in the connection-based method is $ \lceil (3L^2-3)/ 2 \rceil$. Therefore, the theoretical lower bound for $r_c$ is 
\begin{equation}
    r_{\text{min}}(L) = \frac{\left\lceil (3L^2-3)/ 2  \right\rceil}{4L(L-1)},
\end{equation}
so that $\lim_{L \to \infty} r_{\text{min}}(L) = 3/8 = 0.375$. It is natural to ask whether the most efficient connection rule approaches the deterministic rule as the number of choices $k$ increases. Using our largest system size $L = 20$, we note that $r_{\text{min}}(20) = 599/1520 \approx 0.3941$; for this system, varying $k$ and asking how $r_c$ varies, in Fig.~\ref{fig:link_analysis}(a) we use a log-log plot and see that $\log (r_c - r_{\text{min}}) \sim C \log k $ with $C \approx -0.89$, suggesting that $r_c  \to r_{\text{min}}$ as $k \to \infty$ via a simple power law behaviour. 

For the least efficient connection rule, it is easy to see that the overall kirigami pattern is not necessarily rigid if there are two missing connections in any corner tile. Therefore, the upper bound for $r_c$ should be close to 1 in the theoretical deterministic case. Here we again consider $L = 20$ and assess how $1-r_c$ changes with the number of choices $k$. From the log-log plot in Fig.~\ref{fig:link_analysis}(b), we see that $1-r_c$ drops significantly when $k>1$, demonstrating the effectiveness of the power of $k$ choices. However, the result is only marginally improved as we further increase $k$. This suggests that using the least efficient selection rule with $k = 2$ is already sufficient for achieving an explosive percolation in practice.

\begin{figure}[t!]
    \centering
    \includegraphics[width=0.8\linewidth]{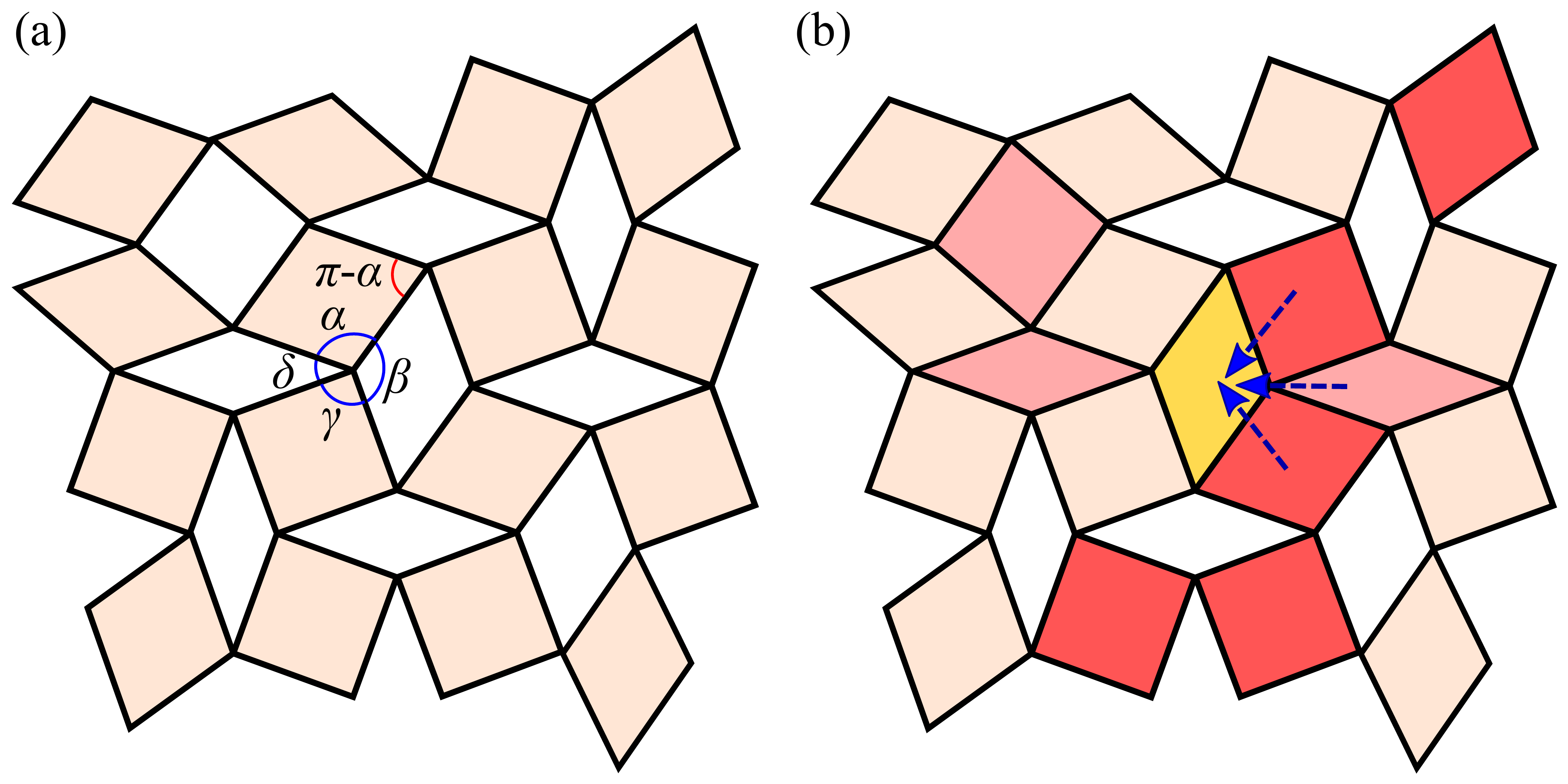}
    \caption{The angle-based method. (a) An illustration of the relationship between different angles. (b) An example pattern with the explicitly rigidified tiles and holes highlighted in red. As indicated by the blue arrows, the hole highlighted in yellow is uniquely determined, i.e. implicitly rigidified.}
    \label{fig:angle_illustration}
\end{figure}

\section{Angle-based method}\label{sect:angle}
Another approach to achieve rigidity percolation is to keep the topology (i.e. the tile connectivity) of a kirigami pattern fixed and change the rigidity of certain components in it. Here we assume that the kirigami pattern has the standard rotating squares topology, and that all tiles have equal side length. Then, the rigidity of a tile can be determined by whether the angles in it are uniquely determined. In particular, from the equal edge length assumption, all tiles and holes must be rhombi and hence the four angles in each of them must be in the form $(\alpha, \pi-\alpha, \alpha, \pi-\alpha)$ for some $\alpha$ (Fig.~\ref{fig:angle_illustration}(a)). Therefore, each of them is uniquely determined if one of its angles is uniquely determined. Now, note that every interior node is shared by four tiles and holes. We have
\begin{equation}
    \alpha+\beta+\gamma+\delta = 2\pi,
\end{equation}
where $\alpha, \beta, \gamma, \delta$ are the angles around an interior node as shown in Fig.~\ref{fig:angle_illustration}(a). This shows that a tile or hole is uniquely determined if the other three tiles and holes have been uniquely determined. Finally, the entire kirigami pattern is rigid if and only if all tiles and holes are uniquely determined.

With this idea in mind, we consider an $L\times L$ kirigami pattern with $L^2$ tiles and $(L-1)^2$ holes in total. In every step, we choose a component (either a tile or a hole) in the pattern that has not been chosen before. If the component is not yet uniquely determined, we prescribe an arbitrary angle in it to fix its geometry. We can then check how many tiles and holes are uniquely determined based on the above angle sum constraints (see Fig.~\ref{fig:angle_illustration}(b) for an example). Note that if the chosen component is already uniquely determined, we simply keep it unchanged and proceed to the next step. The process continues until all $N_{\text{max}} = L^2+(L-1)^2 = 2L^2 -2L +1$ tiles and holes become uniquely determined, i.e. the pattern becomes rigid.
 
Again, the above process of rigidifying a kirigami can be controlled by randomly sampling $k$ components (holes or tiles) and choosing one among them based on a prescribed selection rule in every step. Two selection rules are considered:

\begin{enumerate}[(1)]
\item \textbf{Most efficient angle rule}: Among the $k$ choices of tiles or holes, we choose the one for which fixing its angles gives the maximum total rigid tile and hole count. 

\item \textbf{Least efficient angle rule}: Among the $k$ choices of tiles or holes, we choose the one for which fixing its angles gives the minimum total rigid tile and hole count. 
\end{enumerate}

\begin{figure}[t!]
    \centering
    \includegraphics[width=0.75\linewidth]{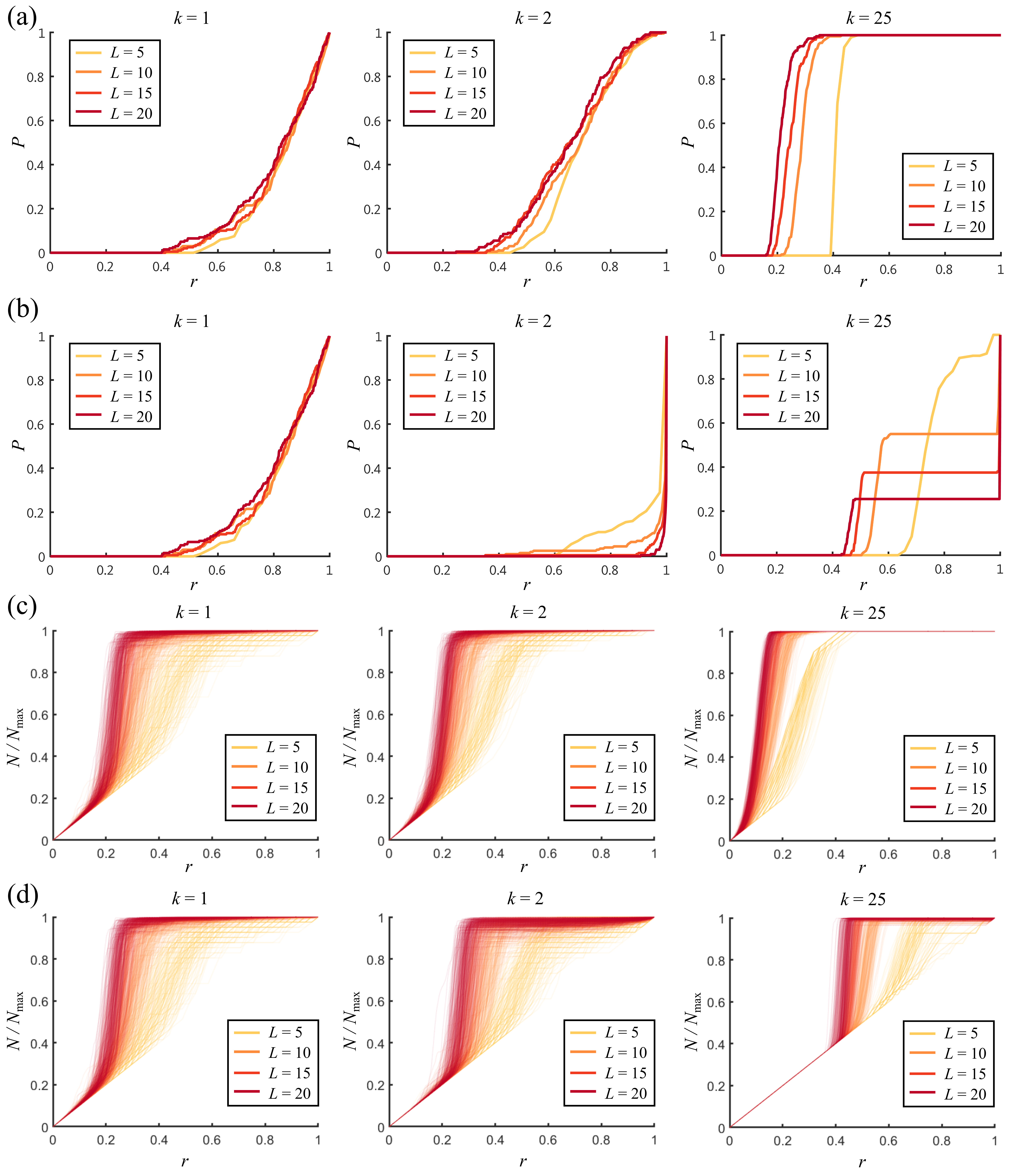}
    \caption{Explosive rigidity percolation achieved using the angle-based method. (a) Using the power of $k$ choices and the most efficient angle rule, we can accelerate the rigidity percolation and achieve a very sharp transition in the probability of getting a rigid pattern $P$ from 0 to 1 at around $r = 0.2$, where $r$ is the percentage of tiles and holes with the rigidity explicitly specified. (b) Using the power of $k$ choices and the least efficient angle rule, we can delay the rigidity percolation and get a very sharp transition in $P$ from 0 to 1 at around $r = 1$. (c) The normalised total rigid tile and hole count obtained using the most efficient angle rule for all 200 simulations for each $L$, where each curve represents one simulation, $N$ is the total rigid count and $N_{\text{max}} = L^2+(L-1)^2 = 2L^2 -2L +1$ is the total number of tiles and holes in an $L\times L$ pattern. (d) The normalised total rigid tile and hole count obtained using the least efficient angle rule for all 200 simulations for each $L$.}
    \label{fig:angle_result}
\end{figure}

To test the performance of the two above rules, here we again consider $L = 5, 10, 15, 20$ and different values of $k$ from $k = 1$ to $k=25$. For each pair $(L, k)$, we perform 200 simulations to prescribe the angles of the tiles and holes gradually and examine the probability of getting a rigid pattern $P$. As shown in Fig.~\ref{fig:angle_result}(a), the rigidity percolation can be significantly accelerated using the most efficient angle rule, with a sharp transition in $P$ occurring at $r = 0.2$ or even smaller. By contrast, it can be observed in Fig.~\ref{fig:angle_result}(b) that the percolation is significantly delayed to occur at around $r = 1$ using the least efficient angle rule. Interestingly, this time the sharp transition in $P$ is not exactly from 0 to 1. Instead, there is first a small increase in $P$ at a smaller $r$, and then a sharp increase to 1 at around $r=1$.

To explain this, we consider how the total rigid tile and hole count $N$ changes as $r$ increases for every simulation. For the most efficient angle rule, one can see that $N/N_{\text{max}}$ increases in a highly nonlinear manner at small $r$ when $k>1$ is used (Fig.~\ref{fig:angle_result}(c)), and all patterns become rigid shortly after the nonlinear increase. By contrast, the least efficient angle rule favors tiles or holes that do not affect the total rigid count. At the early stage of each simulation, tiles and holes that are independent of the other ones will be preferred, and hence $N/N_{\text{max}}$ increases linearly with none of the patterns being rigid (Fig.~\ref{fig:angle_result}(d)), causing a plateau of $P = 0$ for small $r$. However, once $r$ reaches certain values, the new choices of tiles and holes may introduce some dependency of the previous components, which may immediately rigidify a large region or even the entire pattern in some simulations, resulting in $P > 0$. In case the chosen component rigidifies the entire pattern, $N/N_{\text{max}}$ will be equal to 1 throughout the remaining steps. In case the chosen component does not rigidify the entire pattern, the tiles and holes that have not yet been chosen but are implicitly rigidified by the previously chosen ones will now be more preferred, as choosing them will not cause any increase in the total rigid tile and hole count. Therefore, one can see that $N/N_{\text{max}}$ remains slightly below 1 for a large number of simulations in Fig.~\ref{fig:angle_result}(d), causing a second plateau of $P<1$ for a certain period of $r$ in Fig.~\ref{fig:angle_result}(b). Finally, as $r \to 1$, most of the ``redundant'' tiles and holes have already been chosen and explicitly rigidified, and so the remaining choices of the tiles and holes will make the entire pattern rigid and lead to another sharp transition in $P$ from the plateau value to 1.

\begin{figure}[t!]
    \centering
    \includegraphics[width=0.7\linewidth]{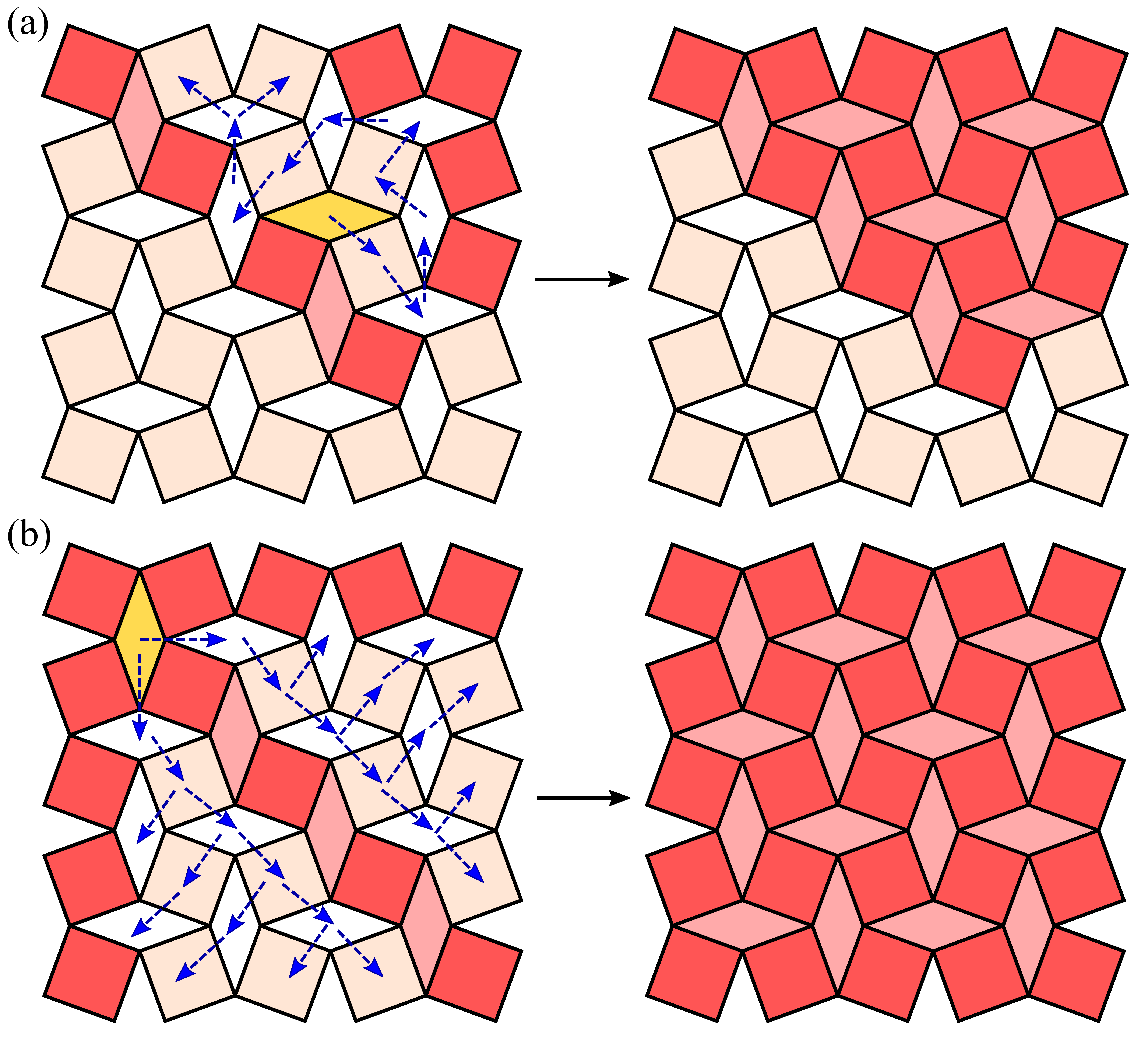}
    \caption{Sharp change in rigidity by the angle-based method. (a) A $5\times 5$ example where the red tiles and holes are already rigidified and the yellow hole is the next chosen component (left). As indicated by the blue arrows, the newly rigidified hole will rigidify other tiles and holes via the angle sum constraint and lead to a large region of rigid tiles and holes (right). (b)~Another $5\times 5$ example pattern with the rigid tiles and holes highlighted in red. After rigidifying the newly chosen hole (highlighted in yellow), the entire pattern will become rigid via a series of angle sum constraints (blue arrows) after exactly $4L-3$ steps.}
    \label{fig:angle_sharp}
\end{figure}

To illustrate the above idea, Fig.~\ref{fig:angle_sharp}(a) shows a $5\times 5$ example with several rigidified tiles and holes (highlighted in red). Once a newly chosen hole (highlighted in yellow) is rigidified, the rigidity of some other tiles and holes will also be enforced by the angle sum constraint (as indicated by the blue arrows), thereby leading to a large rigid region. Those implicitly rigidified tiles and holes will then be preferred under the least efficient angle rule in the subsequent steps, as choosing them does not further increase the total rigid tile and hole count. Fig.~\ref{fig:angle_sharp}(b) shows another $5\times 5$ example with all rigidified tiles and holes (highlighted in red) being independent and hence the overall pattern is achieving minimal rigidity, with the total rigid tile and hole count being exactly $4L-4$ after $4L-4$ steps. However, if the hole highlighted in yellow is rigidified in the next step, it will rigidify all other tiles and holes via the angle sum constraint as indicated by the blue arrows, thereby leading to a rigid pattern after $4L-3$ steps.

We now analyse the two sets of simulation results more quantitatively. For the most efficient angle rule, we again define the critical percentage of explicitly rigidified tiles and holes $r_c$ as the value of $r$ with $P = 1/2$ and study how $r_c$ changes with $k$. Note that in the deterministic case, we have the following result:
\begin{proposition} \label{thm:angle}
The minimum number of steps needed for rigidifying an $L\times L$ pattern (where $L \geq 2$) in the angle-based method is $4L-3$.
\end{proposition}
\noindent {\bf Proof.} We prove the statement by induction.\\
For $L =2$, it is easy to see that the pattern is not uniquely determined unless all $4+1 = 5 = 4\times 2-3$ tiles and holes are fixed.\\
Now, suppose the statement is true for $L$ and consider an $(L+1) \times (L+1)$ pattern. By the induction hypothesis, rigidifying the top left $L \times L$ sub-pattern requires at least $4L-3$ steps. For the remaining $2L+1$ tiles and $2L-1$ holes in the bottommost and rightmost layers, note that the three corner tiles (top right, bottom left and bottom right) have to be rigidified explicitly as they cannot be implicitly rigidified using any of the angle sum constraints. Also, even after rigidifying the $L \times L$ sub-pattern and the three corner tiles, one can see that the hole adjacent to the bottom right corner tile is still not uniquely determined, and hence at least one more step is needed. Therefore, rigidifying any $(L+1) \times (L+1)$ pattern requires at least $4L-3+3+1 = 4L+1 = 4(L+1)-3$ steps, and an explicit example of a rigid pattern with $4(L+1)-3$ steps can be constructed as shown in Fig.~\ref{fig:angle_sharp}(b). This completes the proof. \hfill $\blacksquare$\\

\begin{figure}[t!]
    \centering
    \includegraphics[width=0.85\linewidth]{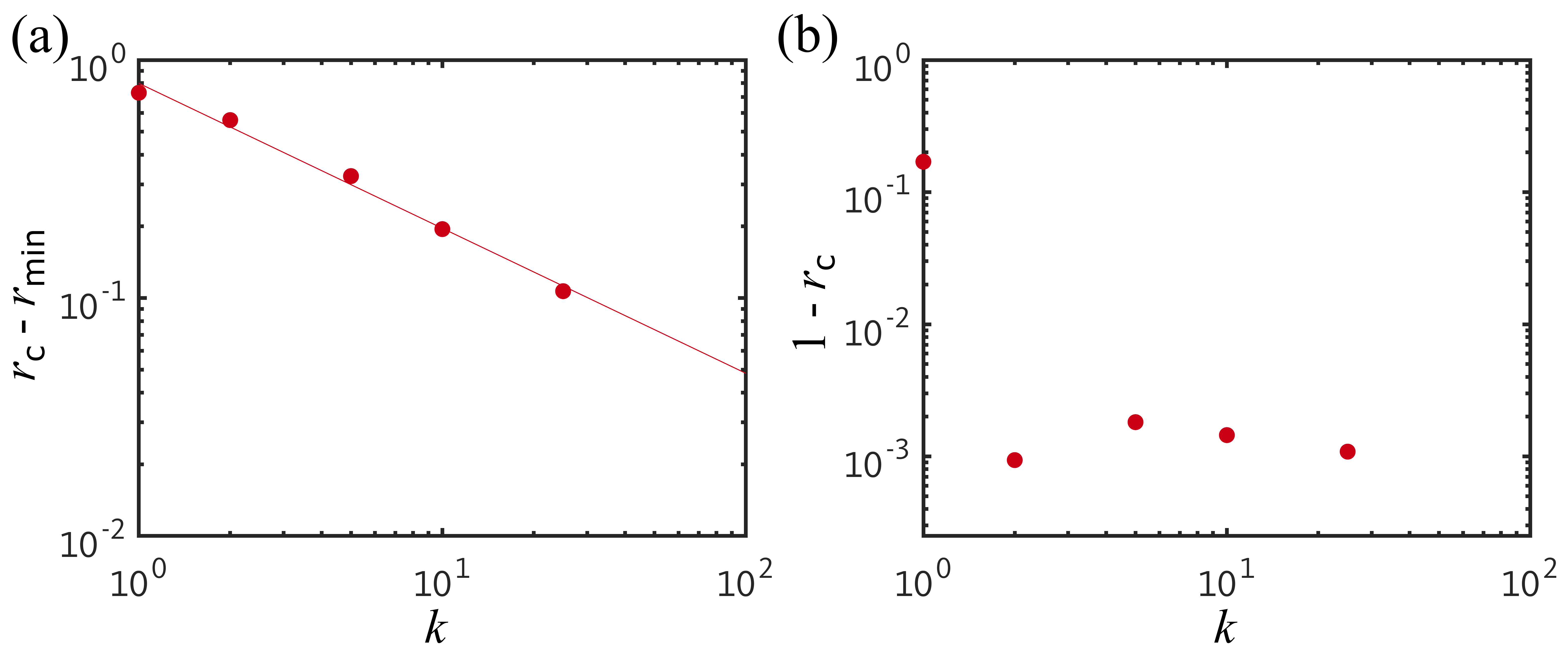}
    \caption{Analyses of different selection rules for the angle-based method. (a) A log-log plot of the number of choices $k$ and the difference $r_c - r_{\text{min}}$ for the most efficient angle rule, with $L = 20$. Each dot corresponds to one choice of $k$ ($k = 1, 2, 5, 10, 25$), and the red line is the best-fit straight line. (b) A log-log plot of the number of choices $k$ and the difference $1 - r_c$ for the least efficient angle rule, with $L = 20$.}
    \label{fig:angle_analysis}
\end{figure}

Consequently, the theoretical lower bound for $r_c$ is
\begin{equation}
    r_{\text{min}}(L) = \frac{4L-3}{2L^2-2L+1}.
\end{equation}
In particular, $\lim_{L \to \infty} r_{\text{min}}(L) = 0$. As shown in the log-log plot for $L = 20$ in Fig.~\ref{fig:angle_analysis}(a), $\log k$ and $\log (r_c - r_{\text{min}})$ form a linear relationship with slope $\approx -0.61$. This suggests that $r_c \to r_{\text{min}}$ as $k \to \infty$, which is close to 0 for large $L$. In other words, we can significantly accelerate the rigidity percolation using the most efficient angle rule. 

For the least efficient angle rule, note that the four corner tiles are always floppy unless explicitly rigidified. Therefore, in theory it takes $2L^2-2L+1$ steps following the least efficient angle rule to ensure that the entire pattern is rigid. In other words, the theoretical upper bound for the critical $r$ is 1. Now, note that because of the presence of the plateau of $P$, it may not be suitable to use $P = 1/2$ to capture the sharp transition. Instead, here we define $r_c$ as the value of $r$ with $P = (1+\text{plateau value})/2$. As shown in the log-log plot for $L = 20$ in Fig.~\ref{fig:angle_analysis}(b), there is a significant decrease in $1-r_c$ as $k > 1$ is used, which again shows the effectiveness of the power of $k$ choices. The change in the difference is much less significant as we further increase $k$, which suggests that using $k = 2$ is already sufficient for deferring the rigidity percolation in practice.

\section{Coordinate-based method}\label{sect:coordinate}
In addition to associating rigidity with whether component (tile or hole) angles are uniquely determined, one may also allow vertex coordinates to be determined. Here, we assume that all tiles are parallelograms. If the vertex coordinates of a component are uniquely determined, the component is considered to be rigid. Rigidity spreads throughout the pattern faster in this formulation because of the extra information the coordinates provide.

\begin{figure}[t]
    \centering
    \includegraphics[width=0.7\textwidth]{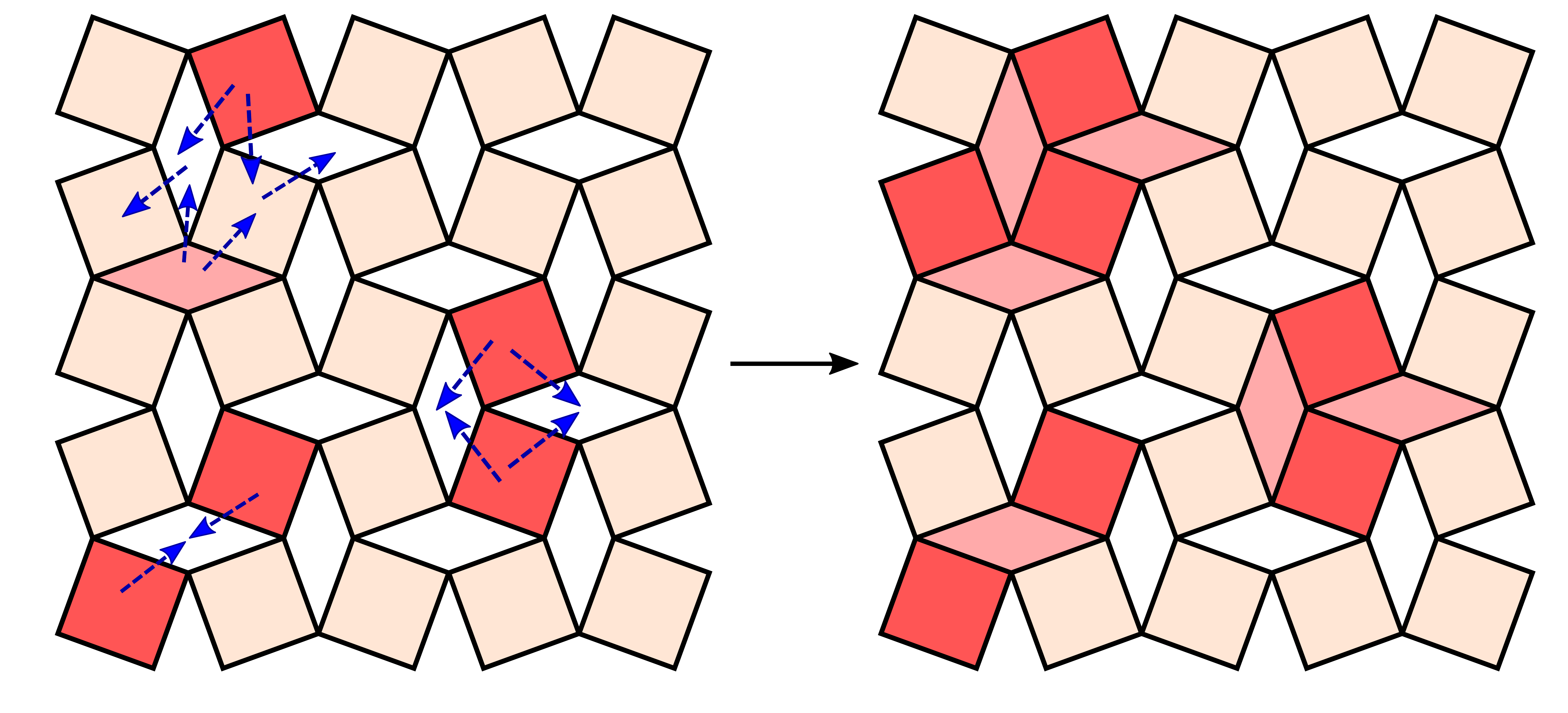}
    \caption{An illustration of the rigidity propagation rule in the coordinate-based method. Rigidifying the tile and hole in the upper left makes an additional four tiles and holes neighboring them rigid. Rigidifying the two red tiles on the right side makes the two neighboring holes rigid. Rigidifying the two opposite red tiles at the bottom left corner makes the hole in between them rigid.}
    \label{fig:coordinate_illustration}
\end{figure}

More specifically, rigidity in the kirigami pattern will propagate via a local rule analogous to the angle sum rule in the angle-based method. If a component $X$ has three vertices which share coordinates with vertices of some rigid components, then component $X$ becomes uniquely determined (see Fig.~\ref{fig:coordinate_illustration}). This is because knowing the coordinates of three vertices uniquely determines the location of the last one if the tile or hole is a parallelogram as we assume.

We simulate the coordinate-based method again by selecting one tile (unlike in the angle-based method, we only select tiles) per round to explicitly rigidify. Tiles that were previously explicitly rigidified are not eligible to be sampled, but tiles that are considered rigid only from the local rule propagation are eligible. To implement explosive percolation, $k$ candidates are sampled and a selection rule is applied to determine which to rigidify. Two selection rules are considered:

\begin{enumerate}[(1)]
\item \textbf{Most efficient coordinate rule}: Among the $k$ choices of tiles, we choose the one for which fixing the vertex coordinates of it gives the maximum total rigid tile and hole count. 

\item \textbf{Least efficient coordinate rule}: Among the $k$ choices of tiles, we choose the one for which fixing the vertex coordinates of it gives the minimum total rigid tile and hole count. 
\end{enumerate}
Note that there are in total $L^2$ tiles in an $L\times L$ pattern and hence the total number of steps is $L^2$. 

\begin{figure}[t!]
    \centering
    \includegraphics[width=0.75\linewidth]{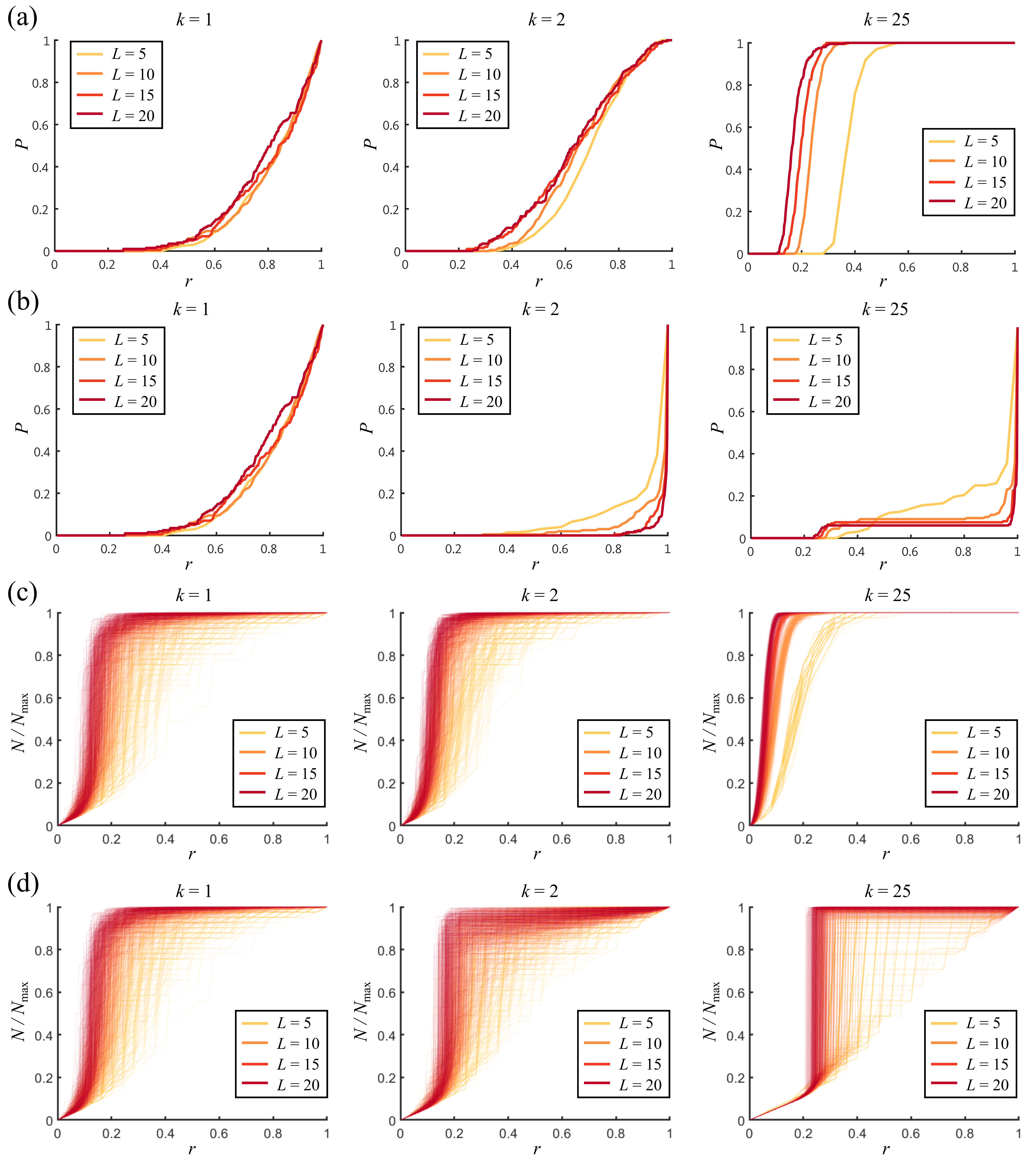}
    \caption{Explosive rigidity percolation achieved using the coordinate-based method. (a) Using the power of $k$ choices and the most efficient coordinate rule, we can accelerate the rigidity percolation, with a sharp transition in the probability of getting a rigid pattern $P$ from 0 to 1 at around $r = 0.15$. (b) Using the power of $k$ choices and the least efficient coordinate rule, we can delay the rigidity percolation, with a sharp transition in the $P$ from 0 to 1 at around $r = 1$. Here $r$ is the percentage of tiles randomly made rigid. (c) The normalised total rigid tile and hole count obtained using the most efficient coordinate rule for all 200 simulations for each $L$, where each curve represents one simulation, $N$ is the total rigid count and $N_{\text{max}} = L^2+(L-1)^2 = 2L^2 -2L +1$ is the total number of tiles and holes in an $L\times L$ pattern. (d) The normalised total rigid tile and hole count obtained using the least efficient coordinate rule for all 200 simulations for each $L$.}
    \label{fig:coordinate_result}
\end{figure}

Again, we consider $L = 5, 10, 15, 20$ and different choices of $k$, and perform 200 simulations for each pair $(L, k)$. Similar to the angle-based method, the most efficient coordinate rule significantly accelerates the rigidity percolation, with a sharp transition at around $r = 0.15$ (Fig.~\ref{fig:coordinate_result}(a)), while the least efficient coordinate rule pushes the percolation back to around $r = 1$ (Fig.~\ref{fig:coordinate_result}(b)). We also consider the normalised total rigid tile and hole count for each simulation. For the most efficient rule, it can be observed in Fig.~\ref{fig:coordinate_result}(c) that the increase in $N/N_{\max}$ is even sharper than that in the angle-based method. By contrast, for the least efficient rule, we can see in Fig.~\ref{fig:coordinate_result}(d) that the delaying effect here is not as significant as that in the angle-based method. The reason is that with the local rule in the coordinate-based method, more tiles and holes are implicitly rigidified in every step.

\begin{figure}[t]
    \centering
    \includegraphics[width=0.75\textwidth]{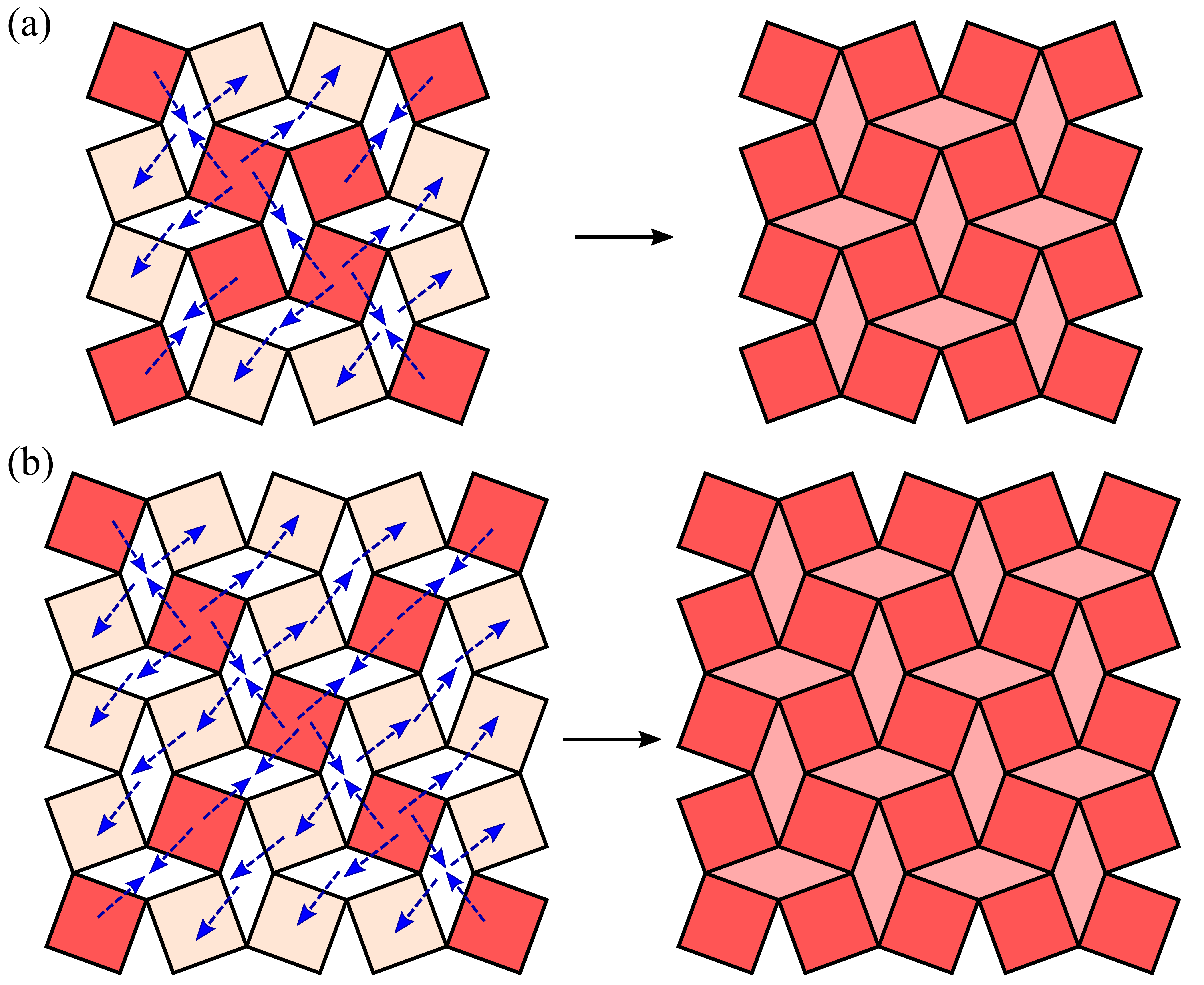}
    \caption{Explicit examples of achieving a rigid pattern in the coordinate-based method using the minimum number of steps. (a) A $4\times 4$ pattern where the $2\times 4 = 8$ red tiles in the left panel are explicitly rigidified. As indicated by the blue arrows, the rigidity will propagate via the local rules, yielding a rigid pattern. (b) A $5\times 5$ pattern where the $2\times 5 -  = 9$ red tiles in the left panel are explicitly rigidified. As indicated by the blue arrows, the rigidity will propagate via the local rules, yielding a rigid pattern.}
    \label{fig:coordinate_sharp}
\end{figure}

To quantitatively analyse the results, we first establish the following deterministic result:
\begin{proposition}
The minimum number of steps needed for rigidifying an $L\times L$ pattern (where $L \geq 2$) in the coordinate-based method is $2L$ if $L$ is even and $2L-1$ if $L$ is odd. 
\end{proposition}
\noindent {\bf Proof.} We prove the statement by induction using a similar argument as in Proposition~\ref{thm:angle}.\\For $L = 2, 3$, note that the four corner tiles in an $L \times L$ pattern have to be rigidified explicitly as they cannot be implicitly rigidified using the local rule. Therefore, we need at least 4 steps. This shows that the statement is true for $L = 2$. For $L = 3$, note that the central tile is still not rigid and hence we need at least one more step, i.e. $5 = 2\times 3 - 1$ steps. It is easy to check that rigidifying the four corner tiles and the central tile indeed gives a rigid pattern. Hence, the statement is also true for $L = 3$.\\
Now, suppose the statement is true for $L$ and consider an $(L+2) \times (L+2)$ pattern. By the induction hypothesis, rigidifying the central $L \times L$ sub-pattern requires at least $2L$ steps (if $L$ is even) or $2L-1$ steps (if $L$ is odd). For the remaining tiles and holes at the boundary layers, again we note that the four corner tiles have to be rigidified explicitly. Therefore, rigidifying any $(L+2) \times (L+2)$ pattern requires at least $2L+4 = 2(L+2)$ steps (if $L$ is even) or $2L-1+4 =2L+3 = 2(L+2)-1$ steps (if $L$ is odd), and explicit examples can be constructed as shown in Fig.~\ref{fig:coordinate_sharp} by rigidifying the two tile diagonals. This completes the proof.\hfill $\blacksquare$\\

Therefore, the theoretical lower bound for $r_c$ is
\begin{equation}
    r_{\text{min}}(L) = \left\{\begin{array}{cl} (2L-1) / L^2 & \text{ if $L$ is odd,}\\ (2L)/L^2 = 2 / L & \text{ if $L$ is even.}\end{array}\right.
\end{equation}
In particular, $\lim_{L \to \infty} r_{\text{min}}(L) = 0$. As for the theoretical upper bound, we have $r_{\text{max}} = 1$ as the corner tiles are always floppy unless explicitly rigidified.

\begin{figure}[t!]
    \centering
    \includegraphics[width=0.8\linewidth]{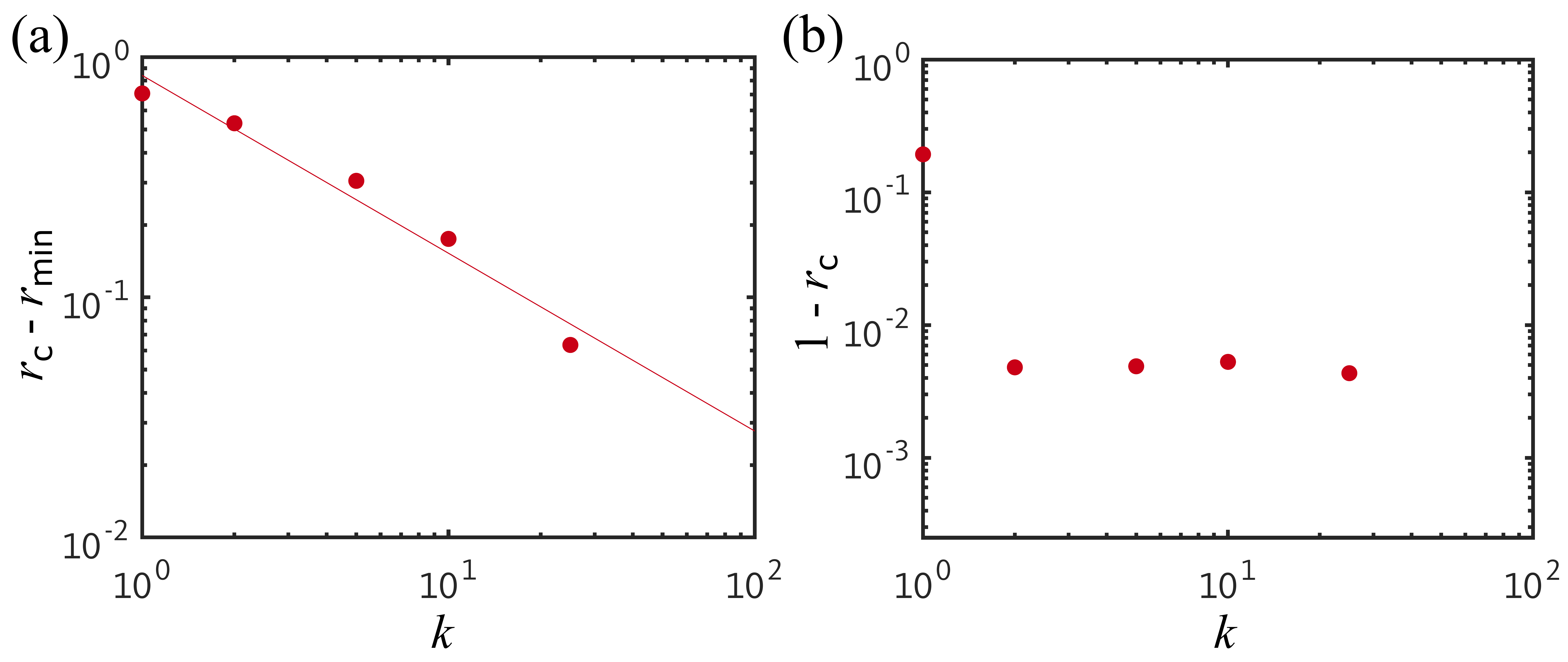}
    \caption{Analyses of different selection rules for the coordinate-based method. (a) A log-log plot of the number of choices $k$ and the difference $r_c - r_{\text{min}}$ for the most efficient coordinate rule, with $L = 20$. Each dot corresponds to one choice of $k$ ($k = 1, 2, 5, 10, 25$), and the red line is the best-fit straight line. (b) A log-log plot of the number of choices $k$ and the difference $1 - r_c$ for the least efficient coordinate rule, with $L = 20$. }
    \label{fig:coordinate_analysis}
\end{figure}

For the most efficient coordinate rule, we define $r_c$ as the critical $r$ with $P = 1/2$. As shown in the log-log plot for $L = 20$ in Fig.~\ref{fig:coordinate_analysis}(a), $\log k$ and $\log(r_c - r_{\text{min}})$ form a linear relationship with slopes $\approx -0.74$, which suggests that $r_c \to r_{\text{min}}$ as $k \to \infty$ and hence the most efficient coordinate rule can effectively accelerate the rigidity percolation. For the least efficient coordinate rule, we follow our approach in the analysis of the angle-based method and define $r_c$ as the value of $r$ with $P = (1+\text{plateau value})/2$. From the log-log plot in Fig.~\ref{fig:coordinate_analysis}(b), we can again see a significant decrease in $1-r_c$ as $k>1$ is used, which implies that the power of $k$ choices can effectively defer the rigidity percolation.

\section{Rigidity of the kirigami pattern with a few components ignored}
\begin{figure}[t!]
    \centering
    \includegraphics[width=\linewidth]{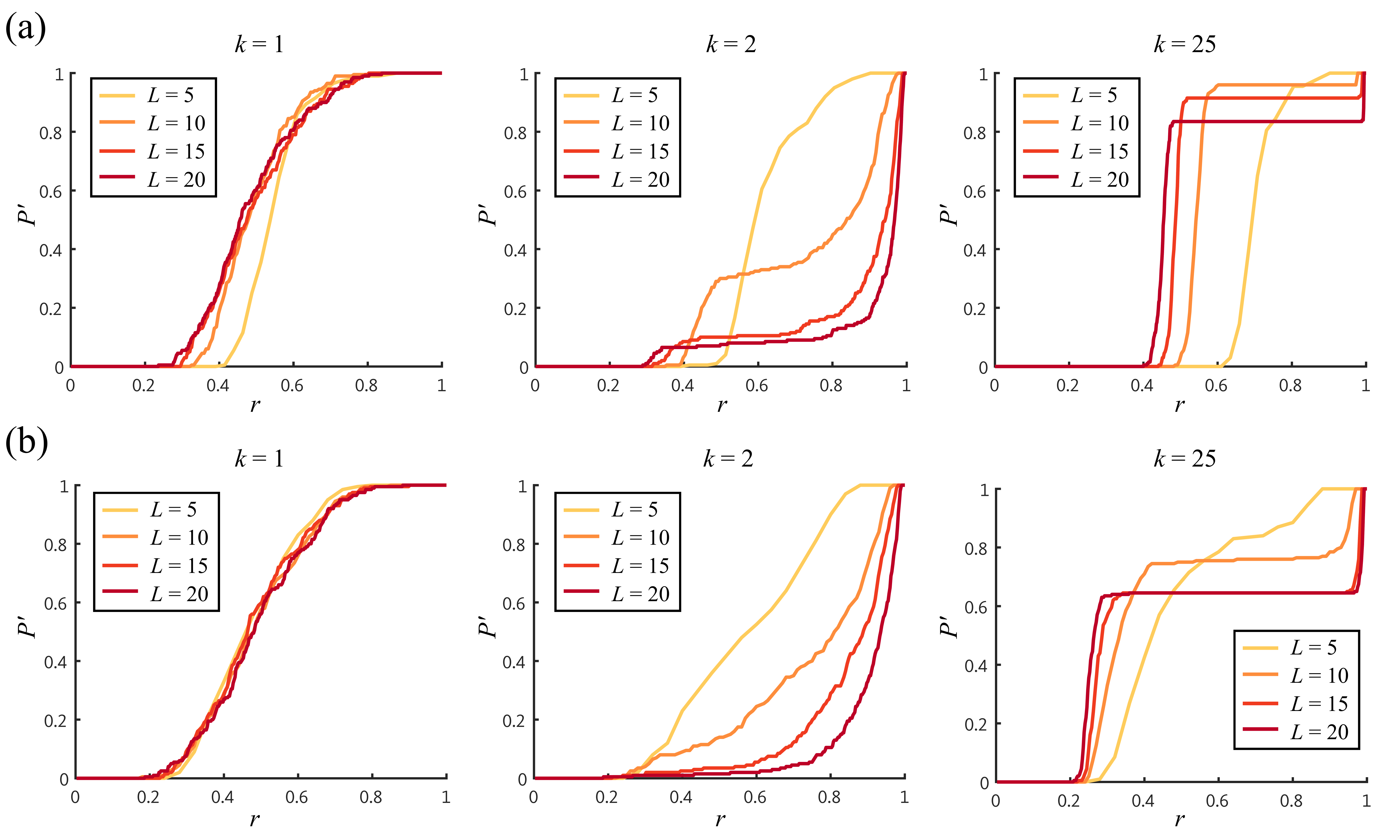}
    \caption{The probability of getting a nearly rigid pattern with up to 4 floppy components using the least efficient rule. (a)~The angle-based method. (b) The coordinate-based method.}
    \label{fig:result_ignoring_corners}
\end{figure}

As shown in the above results, the least efficient rule can significantly delay the occurrence of the sharp transition in the probability of getting a rigid pattern for both the angle-based method and the coordinate-based method. It is noteworthy that in both methods, the kirigami structures are never fully rigid until all the four corner tiles are explicitly rigidified. However, one may consider the floppiness of the corner tiles as an exception and want to focus on the overall rigidity of the structures with a few components ignored. Here, we allow up to 4 components to be floppy for a pattern to be counted as ``nearly rigid'' and recompute the probability of getting a nearly rigid pattern $P'$ in our previously shown simulations using the least efficient rule. As shown in Fig.~\ref{fig:result_ignoring_corners}, the trends of $P'$ look very different from the trends of $P$ in Fig.~\ref{fig:angle_result}(b) and Fig.~\ref{fig:coordinate_result}(b). For instance, for $L = 20$, there is a transition in $P'$ from 0 to 0.8 at about $r = 0.4$ for the angle-based method, which is much sharper than the transition in $P$ from 0 to 0.2. Similarly, there is a transition in $P'$ from 0 to 0.6 at about $r = 0.2$ for the coordinate-based method, which is much sharper than the corresponding transition in $P$ from 0 to about 0.05. This suggests that a large number of patterns are actually rigid except for a few components. The transitions match the explosive jump in the normalised total rigid tile and hole count we observed in Fig.~\ref{fig:angle_result}(d) and Fig.~\ref{fig:coordinate_result}(d). 

\section{Discussion}
In this work, we have shown how explosive rigidity percolation in kirigami can be achieved by changing either the tile connectivity or the rigidity of individual components. Specifically, using the power of $k$ choices and simple selection rules associated with the choice of connection between tiles, changing internal angles or vertex coordinates, we can significantly accelerate or delay the occurrence of the rigidity percolation transition in kirigami. More broadly, our shift in perspective from a purely deterministic or stochastic approach to a hybrid approach may be useful for the design of a wider class of mechanical metamaterials. 

As we have seen, the form of the explosive rigidity percolation is intimately connected to the presences of corners of the quad kirigami patterns. Natural next steps include the study of rigidity percolation in kirigami structures without topological corners, such as patterns created on a disk or an annulus, for which the discrepancy between the transition in $P$ and that in $P'$ might be much smaller. In a different direction, just as  deterministic and stochastic controls of rigidity have also been considered separately in both origami patterns~\cite{chen2019rigidity} and prismatic assemblies~\cite{choi2020control}, it would be natural to extend the current study to analyse and control explosive percolation behaviors in these systems as well.

\bibliographystyle{ieeetr}
\bibliography{kirigami_percolation_bib}

\end{document}